\documentclass[%
  reprint,
superscriptaddress,
 amsmath,amssymb,
 aps,
pra
]{revtex4-2}
\usepackage{empheq}
\usepackage[english]{babel}

\usepackage[letterpaper,top=2cm,bottom=2cm,left=3cm,right=3cm,marginparwidth=1.75cm]{geometry}

\usepackage{braket}
\usepackage{physics}
\usepackage{amsmath}
\usepackage{graphicx}
\usepackage[colorlinks=true, allcolors=blue]{hyperref}
\usepackage{float}
\usepackage{cancel}
\usepackage{ulem}
\usepackage{color}
\usepackage{graphicx,amsmath}%
\usepackage{color,xcolor}
\usepackage{amssymb}
\usepackage{mathrsfs}
\usepackage{subcaption}
\usepackage{upgreek}
\hypersetup{colorlinks=true,citecolor=cyan,linkcolor=blue,urlcolor=blue,pdfstartview=FitH,bookmarksopen=true}
\begin{document}

\title{Incoherent repumping scheme in the $^{88}\text{Sr}^{+}$ five-level manifold}
\author{Valentin Martimort}
\thanks{These authors contributed equally to this work.}
\affiliation{Laboratoire Matériaux et Phénomènes Quantiques, Université Paris Cité, CNRS UMR 7162, 75013 Paris, France}

\author{Sacha Guesne}
\thanks{These authors contributed equally to this work.}
\affiliation{Laboratoire Matériaux et Phénomènes Quantiques, Université Paris Cité, CNRS UMR 7162, 75013 Paris, France}
\affiliation{Crystal Quantum Computing SAS, Incubateur Télécom, 9 Rue Dareau, 75014 Paris, France}

\author{Derwell Drapier}
\thanks{Current address: Physikalisch-Technische Bundesanstalt, Bundesallee 100, 38116 Braunschweig, Germany}
\affiliation{Laboratoire Matériaux et Phénomènes Quantiques, Université Paris Cité, CNRS UMR 7162, 75013 Paris, France}
\affiliation{Laboratoire Kastler Brossel, Sorbonne Université, CNRS,
ENS-Université PSL, Collège de France, 4 place Jussieu, F-75005 Paris, France}

\author{Vincent Tugayé}
\affiliation{Laboratoire Matériaux et Phénomènes Quantiques, Université Paris Cité, CNRS UMR 7162, 75013 Paris, France}

\author{Lilay Gros-Desormeaux}
\affiliation{Laboratoire Matériaux et Phénomènes Quantiques, Université Paris Cité, CNRS UMR 7162, 75013 Paris, France}

\author{Valentin Cambier}
\affiliation{Laboratoire Matériaux et Phénomènes Quantiques, Université Paris Cité, CNRS UMR 7162, 75013 Paris, France}

\author{Albane Douillet}
\affiliation{Laboratoire Matériaux et Phénomènes Quantiques, Université Paris Cité, CNRS UMR 7162, 75013 Paris, France}
\affiliation{Laboratoire Kastler Brossel, Sorbonne Université, CNRS,
ENS-Université PSL, Collège de France, 4 place Jussieu, F-75005 Paris, France}
\affiliation{Université Evry Paris-Saclay, Boulevard François Mitterrand,
F-91000 Evry, France}

\author{Luca Guidoni}
\affiliation{Laboratoire Matériaux et Phénomènes Quantiques, Université Paris Cité, CNRS UMR 7162, 75013 Paris, France}

\author{Jean-Pierre Likforman}
\email{jean-pierre.likforman@u-paris.fr}
\thanks{Corresponding author.}
\affiliation{Laboratoire Matériaux et Phénomènes Quantiques, Université Paris Cité, CNRS UMR 7162, 75013 Paris, France}

\begin{abstract}
Laser-cooled trapped ions are at the heart of modern quantum technologies and their cooling dynamics often deviate from the simplified two-level atom model. Doppler cooling of the $^{88}\text{Sr}^{+}$ ion 
involves several electronic levels and repumping channels that strongly influence fluorescence. In this work, we study a repumping scheme for the $^{88}\text{Sr}^{+}$ ion by combining precision single-ion spectroscopy with comprehensive numerical modeling based on optical Bloch equations including 18 Zeeman sublevels. We show that, although the observed fluorescence spectra retain a Lorentzian lineshape, their width and amplitude cannot be explained by a two-level atom description. Moreover, we find the optimal repumping conditions for maximizing the photon scattering rate.

\end{abstract}
\maketitle

\section{Introduction}

Laser-cooled trapped ions are now widely used in many domains ranging from frequency metrology \cite{Ludlow:2015, Marshall:2025}, precision measurements \cite{Thompson:1990}, quantum simulations \cite{FossFeig:2025} and computing \cite{Cirac:1995, Bruze:2019}.
Atomic ions are  trapped using two families of devices: radio frequency (Paul) traps \cite{Paul:1990} or Penning traps \cite{Dehmelt:1967}. Paul traps can be implemented with different geometries: volume traps (either hyperbolic or linear)  \cite{Ghosh:1995} or surface traps \cite{Chiaverini:2005, Seidelin:2006}.
Many trapped ion applications rely on Doppler laser cooling \cite{Wineland:1975a,Hansch:1975}, either to reach the Doppler limit temperature prior to a specific measurement or as a first step towards lower temperatures achieved through techniques like side-band cooling \cite{Wineland:1975a} or EIT cooling \cite{Morigi:2000}. Moreover, a recent demonstration of the new two-qubit ``smooth'' gate \cite{Hughes:2025} shows that it can achieve high fidelity (error $\sim$ 10$^{-4}$) while relying only on Doppler cooling, without requiring motional ground-state preparation.

In this paper, we performed fluorescence spectroscopy experiments with Doppler-cooled single $^{88}\text{Sr}^{+}$ ions.
This species (as in the case of other alkali-earth ions) exhibits low-lying metastable states requiring the presence of ``repumping'' lasers to keep the electronic population in a closed fluorescence cycle with large scattering rates. 
We developed numerical calculations to describe these experiments.  The model includes five atomic levels (18 Zeeman sublevels) interacting with three laser fields.
We compare experiments and calculations by varying the laser parameters (e.g., intensity and detuning) and measuring the photon scattering rate, which is proportional to the excited-state electronic population.
We were thus able to determine the repumping lasers configurations that optimize the photon scattering rate responsible for Doppler cooling.
We then compare the multi-level system  with the case of a two-level atom, which is often used to model the Doppler laser cooling.

First, we present the atomic system under study, and the relevant interacting laser fields. Then we introduce the model based on the Optical Bloch Equations (OBE) \cite{Allen:1975}.  Next, we describe the experimental setup and protocols. Finally, we present the results obtained both by calculation and experiment, and discuss the comparison with the two-level model.

\section{The $^{88}\text{Sr}^{+}$ five-level manifold}

\begin{figure}[!h] 
\centering
\includegraphics[width=0.4\textwidth]{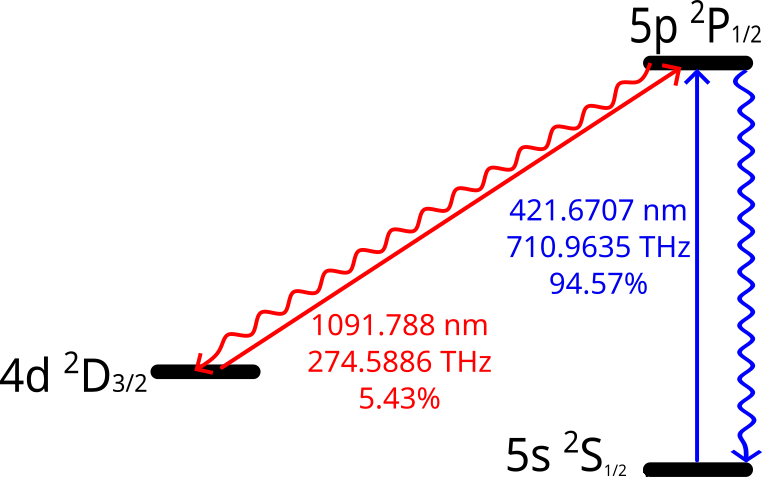}
\caption{The $^{88}\text{Sr}^{+}$ three-level manifold sketch with cooling and repumping lasers in the coherent repumping scheme. A $1092$ nm beam couples the $D_{3/2}$ and $P_{1/2}$ states  with a Rabi frequency noted $\Omega_{1092}/2\pi$. The $422$ nm laser coupling the S and P states is referred as the probe laser of Rabi frequency $\Omega_{422}/2\pi$. Wavy arrows indicate spontaneous emission channels.}
\label{fig:coherent}
\end{figure}

\begin{figure}[h] 
\centering
\includegraphics[width=0.4\textwidth]{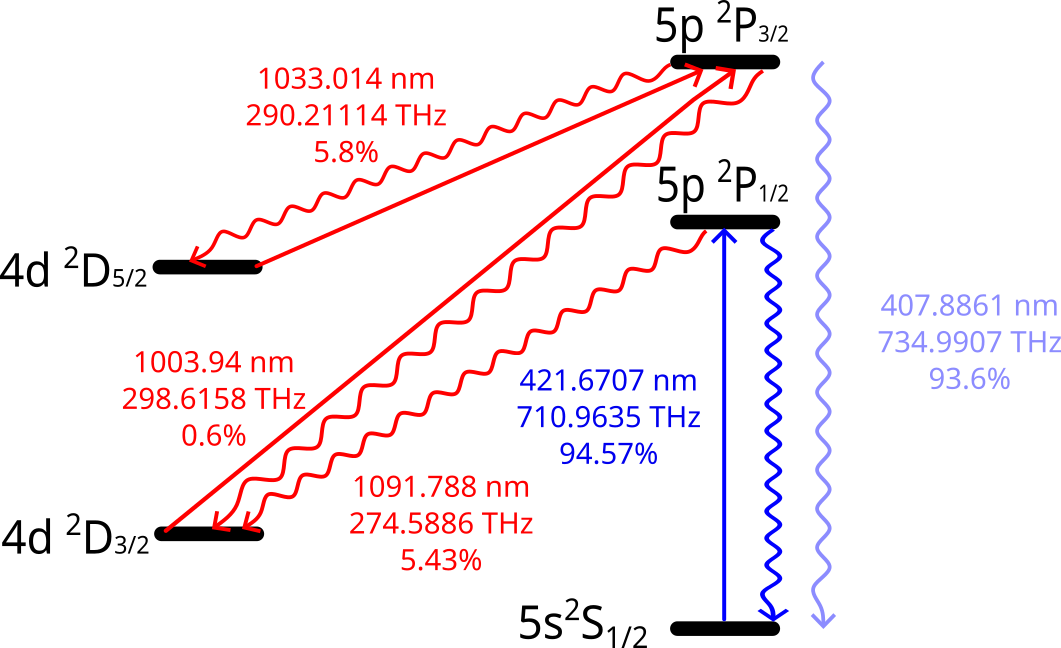}
\caption{The $^{88}\text{Sr}^{+}$ five-level manifold sketch with cooling and repumping lasers in the incoherent repumping scheme. Two infrared lasers at $1003$ nm and $1033$ nm couple respectively the $D_{3/2}$ and $D_{5/2}$ to the $P_{3/2}$ state with Rabi frequencies noted  $\Omega_{1003}/2\pi$ and $\Omega_{1033}/2\pi$ respectively. Wavy arrows indicate spontaneous emission channels}
\label{fig:incoherent}
\end{figure}

The relevant level structure of $^{88}\text{Sr}^{+}$ ion, exploited for Doppler cooling, is shown in Fig. \ref{fig:coherent} and \ref{fig:incoherent}.
We use a $422$ nm beam near resonant with the $S_{1/2}-P_{1/2}$  transition as the ``cooling'' laser.
After the absorption of a $422$ nm photon, the system can relax either back to the fundamental  $S_{1/2}$ state or to the  $D_{3/2}$ metastable state with a probability of 0.056  \cite{Likforman:2016}.
Since the $D_{3/2}$ lifetime is about 400~ms  \cite{Biemont:2000}, a repumping scheme that brings back the electronic population towards the cooling cycle is needed.
Typically, strontium ions are Doppler-cooled using one of the two different repumping strategies, shown in Fig. \ref{fig:coherent} and \ref{fig:incoherent}.
The ``coherent'' repumping scheme  (Fig. \ref{fig:coherent}) uses a single repumping laser that couples the $D_{3/2}$ state to the $P_{1/2}$ state.
In this case, dark resonances appear when the cooling laser, coupling the $S_{1/2}$ state to the $P_{1/2}$ state, has the same frequency detuning as the repumping laser \cite{Arimondo:1976}.

 To avoid dark states, a second repumping scheme \cite{Allcock:2010}, shown in Fig.~\ref{fig:incoherent}, is often used (``incoherent'' repumping scheme).

\section{Motivation}

\begin{figure}[!h]
    \centering
    \includegraphics[width=1\linewidth]{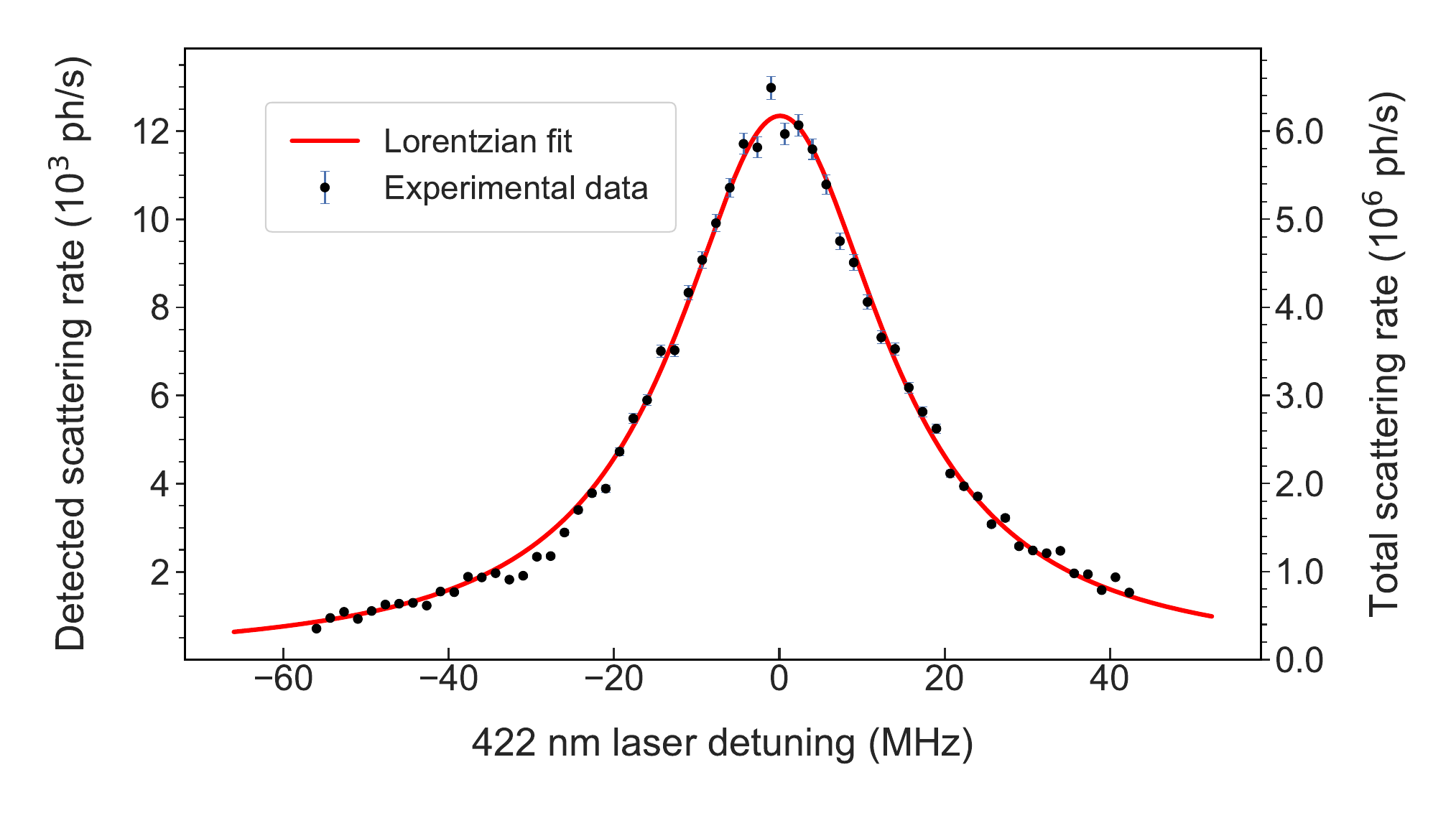}
    \caption{Example of a fluorescence spectrum obtained by varying the 422 nm laser frequency.  At resonance, the detected scattering rate is 12000 photons per second. The lorentzian full width at half-maximum is 30 MHz. The collection efficiency of our setup is $2\times10^{-3}$. The total scattering rate at resonance for this single ion is then $6.5\times10^{6}$ photons/s
    }
    \label{fig:spectrumb}
\end{figure}

This work was motivated by precision spectroscopic measurements with single laser-cooled $^{88}\text{Sr}^{+}$ ions. The implementation of fast sequential acquisition techniques enables the recording of fluorescence spectra free from artifacts  induced by the mechanical action of the probe laser on a single ion (such as Doppler heating) \cite{Gardner:2014, Meir:2014}.
At the same time, this approach provides a powerful tool for determining the collection efficiency of a photon-counting apparatus with a precision better than 1\% \cite{Ramm:2013, Likforman:2016}.
Therefore, by scanning the frequency of a probe laser around the $S_{1/2}-P_{1/2}$  resonance and using the incoherent repumping scheme, we obtain a spectrum such as the one shown in Fig. \ref{fig:spectrumb} where the vertical scale can be corrected using the measured collection efficiency.

As expected, the lineshape of the spectrum in Fig. \ref{fig:spectrumb} is purely Lorentzian as in the case of a two-level atom. 

However, assuming a natural linewidth of $\Gamma/2\pi$=21.5~MHz ($\tau$ = 7.39(7) ns) \cite{Pinnington:1995}, associated with the $P_{1/2}$ state lifetime, the amplitude and linewidth of the experimental spectra do not match the two-level atom model. 
  
In the two-level atom model, the scattering rate is given by the expression:

\[R_{scattering }= \Gamma \frac{s/2}{1 + s + 4\delta^2/ \Gamma^2}\]
 
With $R_{scattering}$ the scattering rate, $\delta$ the detuning, $\delta = (\omega - \omega_{0})/2\pi$ with  $\omega/2\pi$ and $\omega_{0}/2\pi$, respectively, the laser frequency and the ion resonant frequency. s is the saturation parameter, $s = \frac{I}{I_{sat}} = 2  \frac{\Omega^2}{\Gamma^2}$, with $I$ the laser intensity (in W/cm$^2$), $I_{sat}$ the saturation intensity \cite{Foot:2004}, $\Omega/2\pi$  is the Rabi frequency.

The maximum scattering rate at resonance $S$ is related to the power broadened full-width-at-half-maximum $\Gamma_{broad}=\Gamma \sqrt{1+s}$ by : $$S = \dfrac{\Gamma_{broad}}{2}{s}{(1 + s)^{-3/2}}$$

Using the experimental spectrum shown in  Fig. \ref{fig:spectrumb}, we can extract the saturation parameter $s$ from the scattering rate at resonance and we find $s=$~0.14.
Injecting this value into the expression of the linewidth, we find  $22~ \text{MHz}$, far from the observed experimental width of $30 \,\text{MHz}$.
In the case of a two-level atom, a probe laser intensity corresponding to such a broadened spectrum would be associated to a maximum scattering rate of  $33\times10^{6}$ photons/s to be compared to the observed $6.5\times10^{6}$ photons/s.
One might argue that, in the presence of a magnetic field, the Zeeman sublevels could induce additional broadening for a linearly polarized probe beam. However, under the experimental conditions of  Fig. \ref{fig:spectrumb}, the magnetic field applied along the k vector has an amplitude of B = 0.045 mT, resulting in a negligible linewidth broadening of 60 kHz. 

This example shows that in spite of an observed Lorentzian lineshape for the fluorescence spectra, the two-level approximation does not necessarily hold within the incoherent repumping scheme. 

The two-level atom model is often used, for example, in the Doppler-recooling technique  \cite{Wesenberg:2007} to characterize the heating rate of an ion trap.
The Doppler recooling technique has been applied to ions in which metastable states are present \cite{Allcock:2010, Sikorsky:2017}. However, the five-level model that we describe may be useful to understand in which conditions this technique can be exploited.

\section{Calculations}

To fully describe the atom-laser interaction in the case of incoherent repumping of $^{88}\text{Sr}^{+}$, we developed a model based on OBE \cite{Arecchi:1965, Allen:1975}, well adapted to estimate mean values of physical quantities \cite{Grynberg:2010}.
Our approach is similar to that of Janacek and co-workers \cite{Janacek:2018}
consisting in a computer-based solving of the problem that includes the Zeeman sublevels. Since the experimental data are obtained by measuring the ion fluorescence, we focused on calculating all the populations of the five-level manifold, as a function of the lasers parameters i.e. frequencies, intensities and polarizations. This approach is well suited to model fluorescence measurements since it involves averaging on many photons and many experimental realizations. 

We also used the standard dipolar interaction approximation and the rotating wave approximation. The laser fields are treated classically and their profiles are considered Gaussian. Spontaneous emission and decoherence are also treated classically using the decay rates of each state, with best known values taken from the NIST databases \cite{NIST:2025}.

The ions are immersed a magnetic field and all the 18 Zeeman sublevels of the five-level system are included in the model. To our knowledge, such a large number of levels is rarely considered for calculating populations and coherences of ``real'' atoms \cite{Maguire:2006, Janacek:2015}. 

In the following, we consider three laser fields that are all linearly polarized and propagate along the direction of the magnetic field, as in the experiment.

\begin{figure}[H] 
    \centering
    \includegraphics[width=1\linewidth]{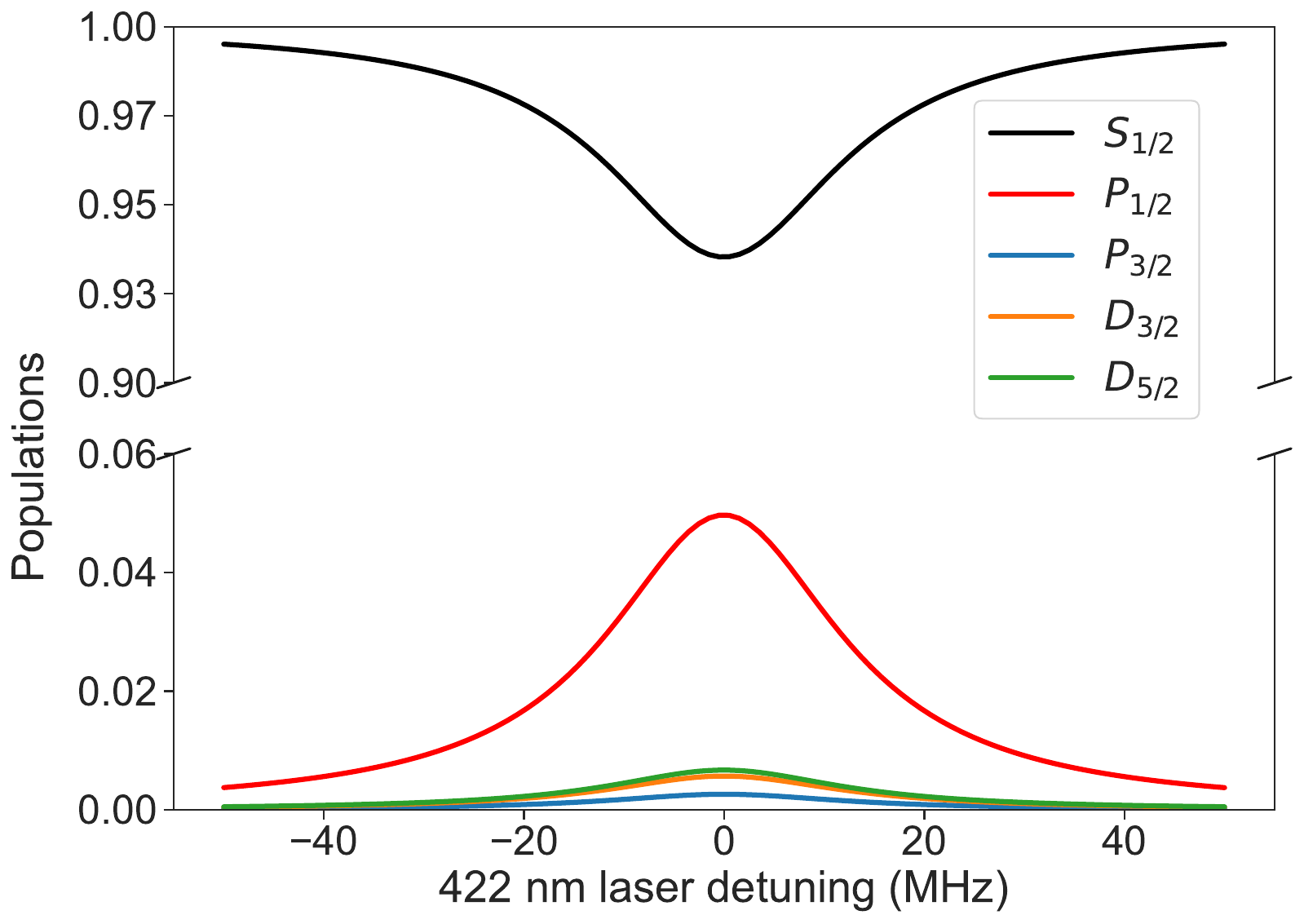}
    \caption{Electronic states populations calculated with the following parameters :  $\Omega_{422}/2\pi = 11\,\text{MHz}$, $\Omega_{1003}/2\pi = 150\,\text{MHz}$, $\Omega_{1033}/2\pi = 250\,\text{MHz}$. The 1003 nm laser is at resonance, while the 1033 nm laser is +400 MHz detuned from resonance. The magnetic field amplitude is $4 \times 10^{-4}$ T.  \color{black}}
    \label{fig:populations}
\end{figure}

The numerical calculations are performed using standard python libraries such as sympy (e.g. to calculate the Clebsch-Gordan coefficients, describing the angular momentum coupling between two states).
We calculated the steady-state populations of $^{88}\text{Sr}^{+}$ at rest both in the incoherent and coherent repumping schemes. More details can be found in  \cite{Tugaye:2020} and in the Appendix.
We perform the calculation of fluorescence spectra such as the one of Fig. \ref{fig:spectrumb} obtained in the incoherent repumping scheme.  Intensities and frequencies of the three lasers  and the magnetic field amplitude are fixed, based on typical experimental conditions (see Fig. \ref{fig:populations} caption).
The populations of the five electronic states summed over the Zeeman sublevels are shown in Fig. \ref{fig:populations}.

\vspace{3cm} 
\begin{figure}[H]

    \begin{subfigure}{0.5\textwidth}
        \centering
        \includegraphics[width=0.8\linewidth]{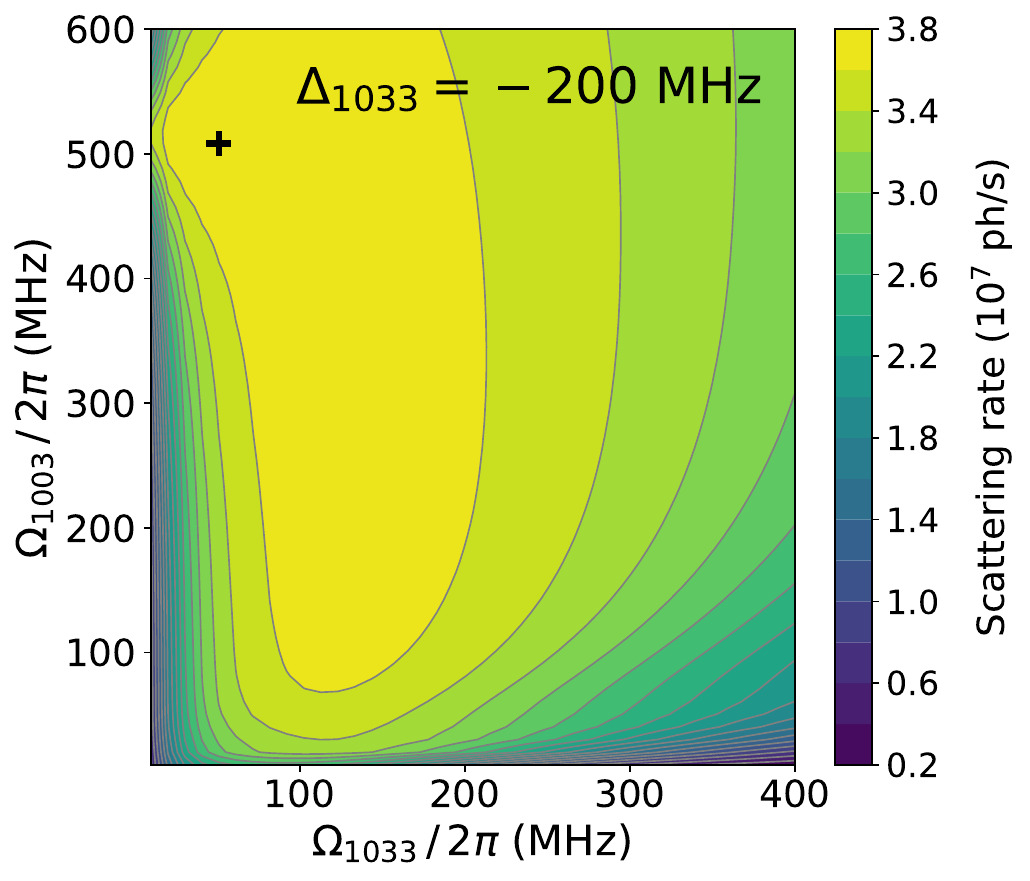} 
        \caption{}
        \label{fig:subim1}
    \end{subfigure}
    
    \begin{subfigure}{0.5\textwidth}
        \centering
        \includegraphics[width=0.8\linewidth]{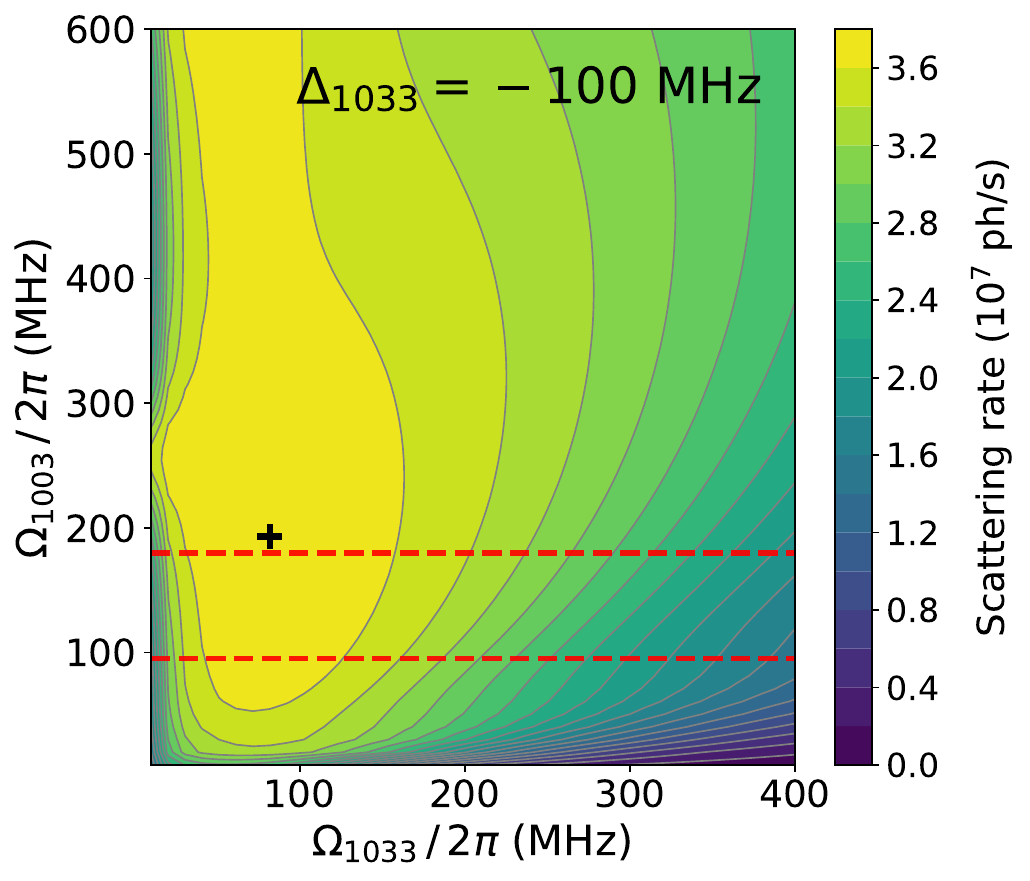}
        \caption{}
        \label{fig:subim2}
    \end{subfigure}

    \begin{subfigure}{0.5\textwidth}
        \centering
        \includegraphics[width=0.8\linewidth]{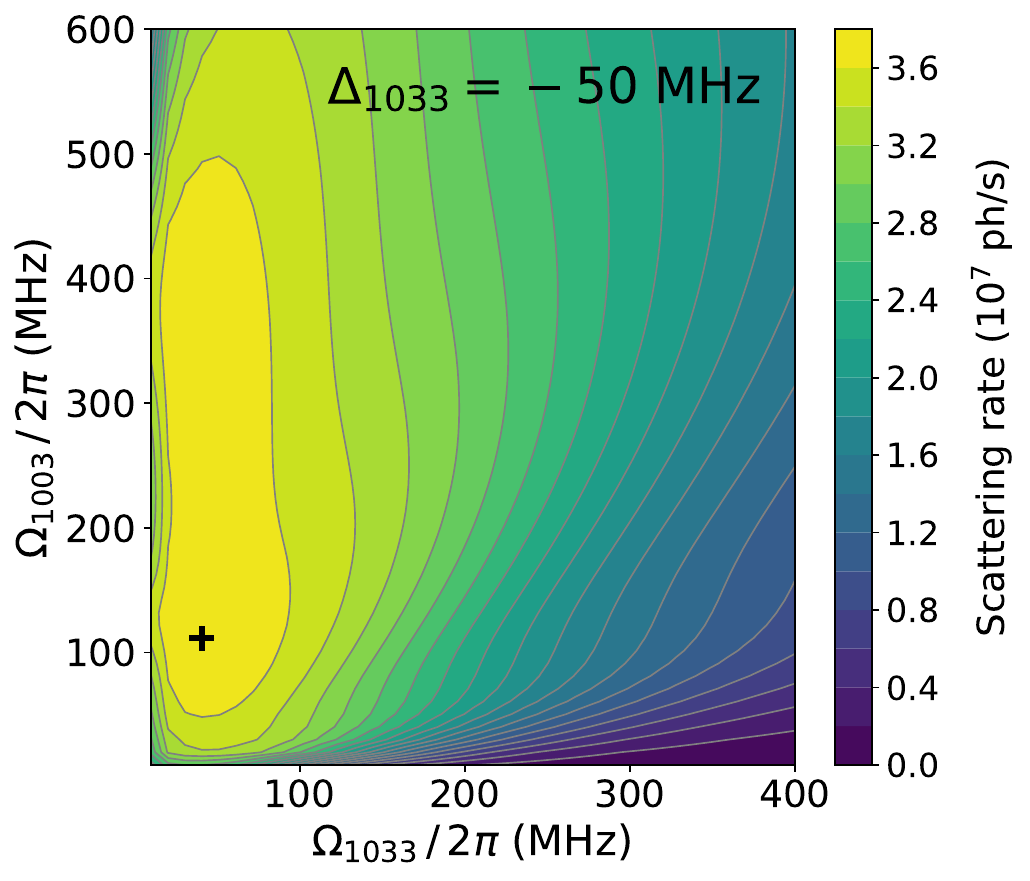}
        \caption{}
        \label{fig:subim3}
    \end{subfigure}

    \caption{Scattering rate  of the 422 nm photons as a function of the repumping lasers Rabi frequencies for a given 422 nm Rabi frequency $\Omega_{422}/2\pi = 37\,\text{MHz}$ and three different 1033 nm laser frequency detunings : (a) $\Delta_{1033} = -200\,\text{MHz}$, (b) $\Delta_{1033} = -100\,\text{MHz}$, (c) $\Delta_{1033} = -50\,\text{MHz}$. The 1003 nm laser is at resonance. The magnetic field amplitude is such that $ B \times \mu_B/h  = 3.6\,\text{MHz}$. The two dotted red lines in (b) correspond to measured two cross-sections that will be presented in Fig. \ref{fig:img1} and Fig. \ref{fig:img2}. For each plot, the position of the maximum value of the scattering rate is indicated by a black cross. These positions,  noted as follows ($\Omega_{1033}/2\pi$ , $\Omega_{1003}/2\pi$) in MHz units are : (51, 508) for (a), (82, 193) for (b) and (41, 112) for (c).}
    \label{fig:map2D}
\end{figure}

\newpage
\noindent
\begin{minipage}{\columnwidth}

\noindent\hspace*{8pt} The calculated fluorescence spectrum (Fig. \ref{fig:Fluo_11}) is the product of the natural linewidth $\Gamma$ times the branching ratio to the $S_{1/2}$ state (0.946 \cite{Likforman:2016}) times the $P_{1/2}$ population.

\begin{figure}[H] 
    \centering
    \includegraphics[width=1\linewidth]{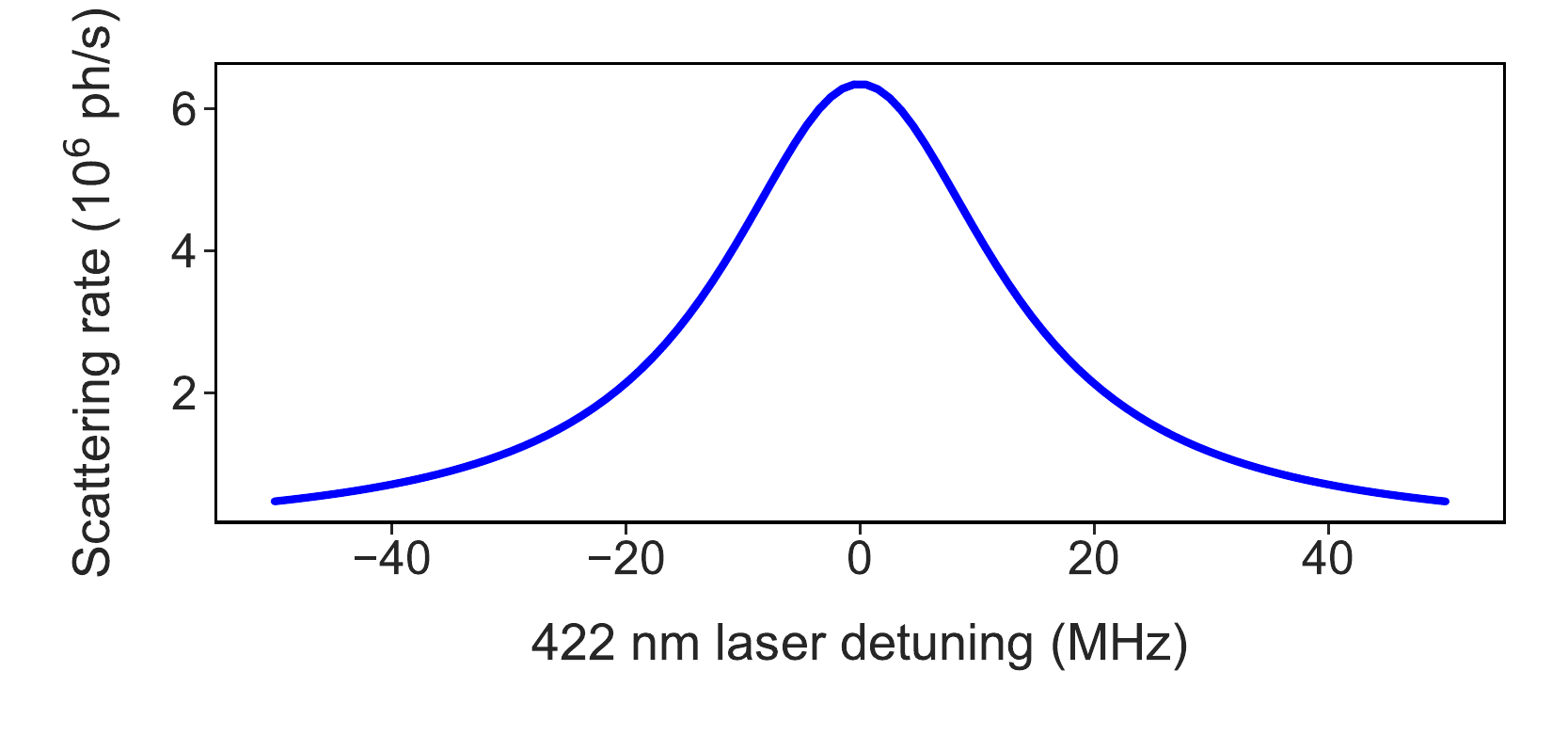}
    \caption{Fluorescence spectrum calculated with the following parameters :  $\Omega_{422}/2\pi = 11\,\text{MHz}$, $\Omega_{1003}/2\pi = 150\,\text{MHz}$, $\Omega_{1033}/2\pi = 250\,\text{MHz}$. The 1003 nm laser is at resonance, while the 1033 nm laser is +400 MHz detuned from resonance. The magnetic field amplitude is $4 \times 10^{-4}$ T.\color{black} }
    \label{fig:Fluo_11}
\end{figure}

\noindent\hspace{8pt} The spectrum of Fig. \ref{fig:Fluo_11} has a Lorentzian shape and is in good agreement with the experimental spectrum of Fig. \ref{fig:spectrumb}, with a maximum scattering rate of $6.3\times10^6$ photons/s and a width of about 30 MHz. 
\vspace{0.5cm}

\noindent\hspace{8pt} To explore different repumping conditions, we vary the laser parameters.  In Fig. \ref{fig:map2D}, we calculate the maximum scattering rate (probe laser at resonance) for various repumping Rabi frequencies and frequency detunings.
\vspace{0.5cm}

\noindent\hspace{8pt} The calculations of  Fig. \ref{fig:map2D} show the somewhat counter-intuitive result that the maximum scattering rate is not obtained with the maximum intensities of the repumping lasers.  The positions of the maximum scattering rates on the 2D-plots depend on the  repumping lasers detunings. This feature is visible in Fig. \ref{fig:map2D} in the case of the $1033$ nm laser detuning.
\vspace{0.5cm}

\noindent\hspace{8pt} In the following section we  describe single-ion experiments.
\end{minipage}

\section{Experimental methods}

\begin{figure}[h] 
    \centering
       \includegraphics[width=0.9\linewidth]{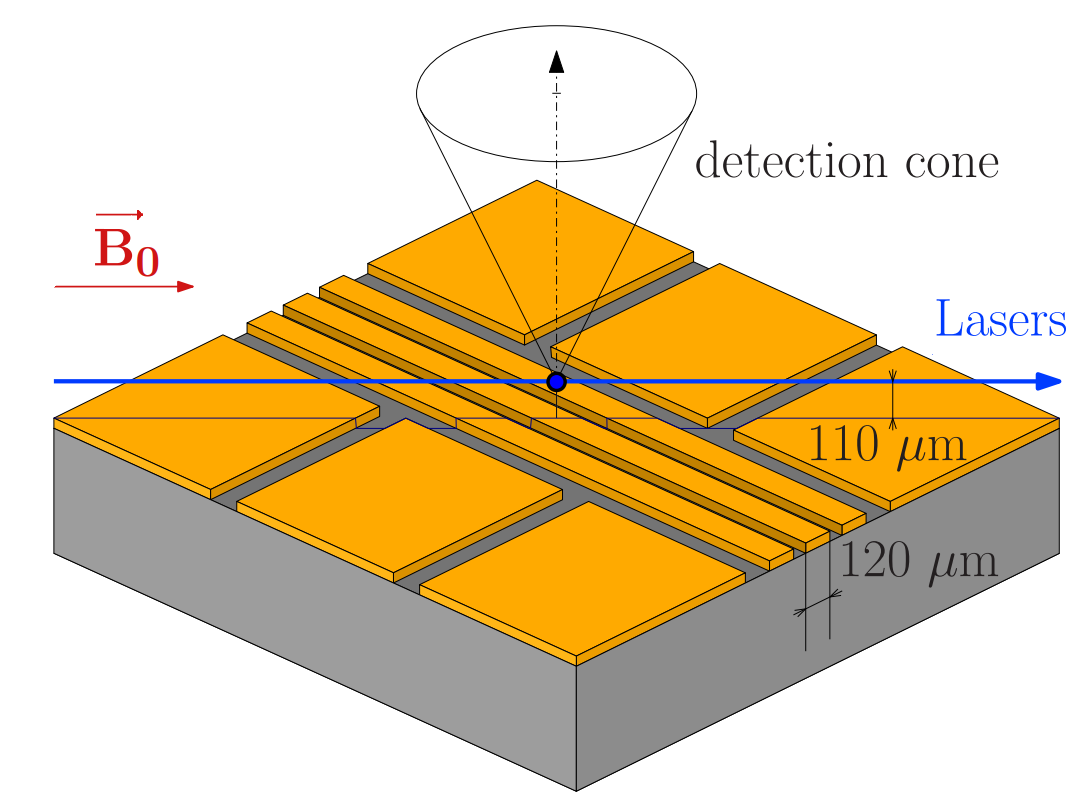}
    \caption{Schematic view of the surface ion trap. The cooling, repumping, and photoionizing laser beams propagate parallel to the trap surface aligned at 45° with respect to the trap axis. A magnetic field $\Vec{B_0}$ with an amplitude of the order of 1 × $10^{-4}$ T defines a quantization axis parallel to the wave vectors of the laser beams.}
    \label{fig:trap_capture}
\end{figure}

The experimental setup is similar to the one described in \cite{Likforman:2016}.  A single ion is Doppler-cooled and trapped in a symmetric five-wire surface trap \cite{Chiaverini:2005} with a nominal ion-surface distance $d = 110~\mu\mathrm{m}$.  The cooling laser at 422 nm and the repumping lasers at 1003, 1033 and 1092 nm are co-linear, linearly polarized and propagating along the direction of the controlled magnetic field, as shown in Fig. \ref{fig:trap_capture}. We measure the ion fluorescence intensity as a function of the repumping lasers intensities using the above mentioned fast sequential acquisition techniques. Considering an ion at rest, its fluorescence intensity depends on the 422 nm probe laser intensity and detuning, the repumping lasers intensities and detunings and on the amplitude of the magnetic field. In the following, we describe how we determine the experimental parameters.

\begin{figure}[h] 
    \centering
       \includegraphics[width=1\linewidth]{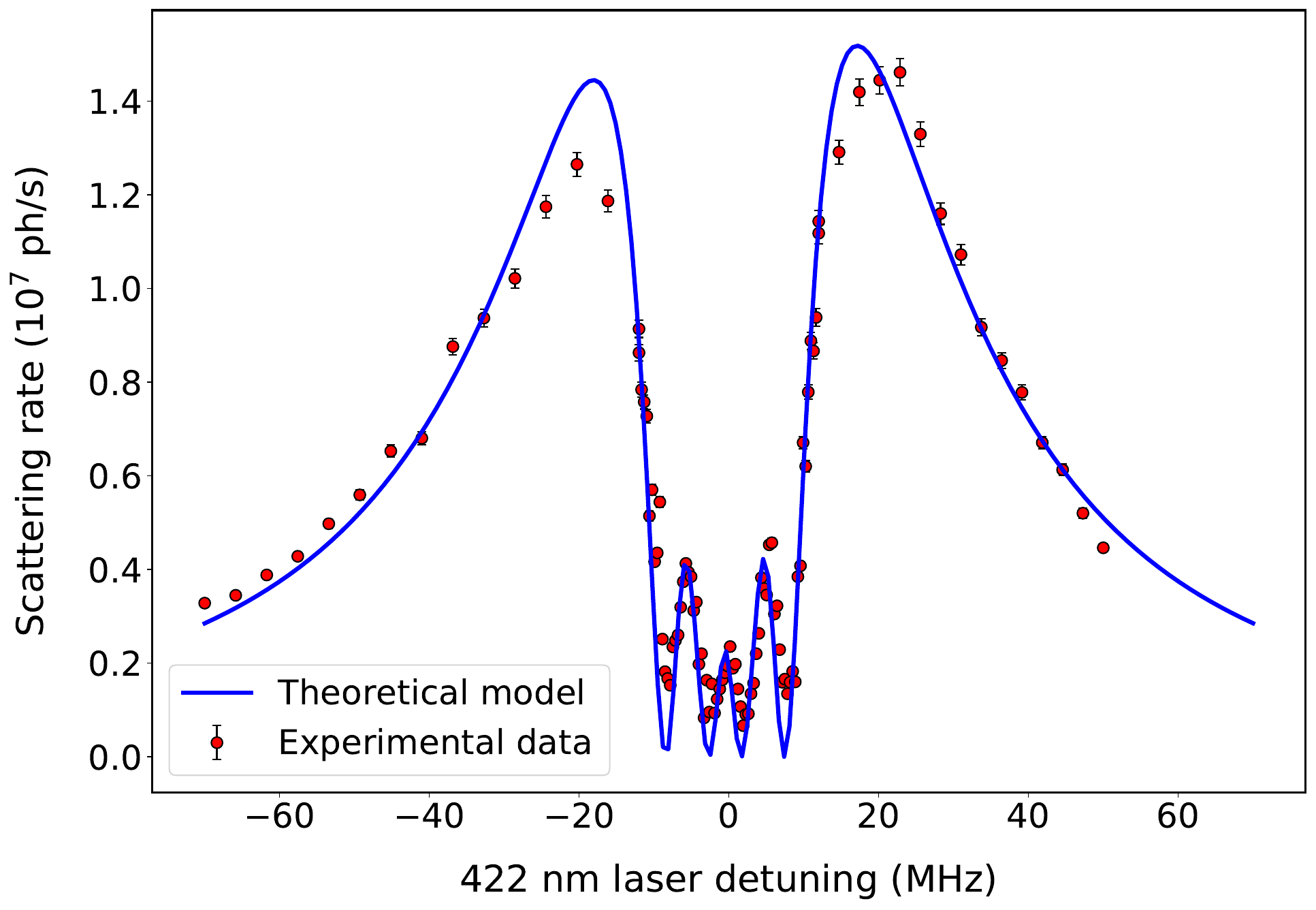}
    \caption{Fluorescence spectrum using the coherent repumping scheme. The blue line corresponds to the above described model with the following parameters :  $1092$ nm laser frequency detuning $\Delta_{1092} = 0 \,\text{MHz}$ ,  $\Omega_{422}/2\pi = 37\,\text{MHz}$,  $\Omega_{1092}/2\pi =9\,\text{MHz}$ and $ B \times \mu_B/h  = 3.6\,\text{MHz}$. The collection efficiency is $ 4.6 \times 10^{-3}$.}
    \label{fig:spectrumIR1}
\end{figure}

In order to measure the amplitude of the magnetic field as well as the 422 nm probe laser intensity, we acquire a fluorescence spectrum using the coherent repumping scheme (Fig. \ref{fig:coherent}).  A typical  spectrum is shown in  Fig. \ref{fig:spectrumIR1}. 

The envelope of the spectrum depends on the 422 nm probe laser intensity and the position of the spectral dips depends on the $1092$ nm laser frequency detuning with respect to the $P_{1/2}$ state. Each dip corresponds to a dark state formed with the Zeeman sublevels of the $S_{1/2}-P_{1/2}-D_{3/2}$ $\Lambda$ system.

The frequency spacings of the four dark resonances are directly proportional to the magnetic field amplitude. In practice, to resolve the four dark resonances, we need to operate in a power regime for both lasers where the power broadening of the spectral lines is smaller than the energy spacing between the Zeeman sublevels. Using the model described in the previous section and the experimental spectrum (Fig. \ref{fig:spectrumIR1}), we can determine the 422 nm probe laser intensity and the amplitude of the magnetic field along the propagation direction such that $ B \times \mu_B/h  = 3.6\,\text{MHz}$ with $\mu_B $ the Bohr magneton and $h$ the Planck constant.

We now describe how we measured the repumping lasers intensities. We measured the laser power of each beam using a calibrated Ge photodetector (818-ST2-IR, Newport). The power range explored for the 1033 nm laser, extends from 20 nW to 500 µW, and for the 1003 nm laser from 50 to 450 µW . To measure the beam's size at the ion position, we sample the beam using a pelicular plate. This plate is positioned in between the focusing lens and the entrance viewport of the vacuum chamber. The sampled beam profile is recorded using a CCD camera. Both the ion and the camera's sensor are equidistant from the focusing lens. To ensure equal optical pathways we introduced a viewport in the sampled beam path. 

Experimental Rabi frequencies are calculated from the laser power and beam profile using the following formula \cite{Foot:2004} :

\[\Omega/ 2\pi= \frac{\mu}{h} \sqrt{\frac{2I}{\epsilon_0 c}}
\]
with h, the Planck constant, $\mu$ the transition dipole moment, $\epsilon_0 $ the vacuum permittivity, c the light velocity in vacuum and I the laser intensity. I is expressed as function of the laser average power P and beam surface S as  $I = 2P/S =  2P/\pi w_xw_y$, with $w_x$ and $w_y$ the beam half-width at $1/e^2$ along eigen axes. The beam surfaces are 1350 and 2200 µm$^2$, for the 1003 nm and 1033 nm lasers, respectively.

The infrared lasers are intensity-stabilized using back-action on an acousto-optic modulator (AOM). Laser power adjustments are implemented by changing the setpoint of the intensity stabilization servo-loop. 

\begin{figure}[h!] 
    \centering
    \includegraphics[height = 0.3\textwidth]{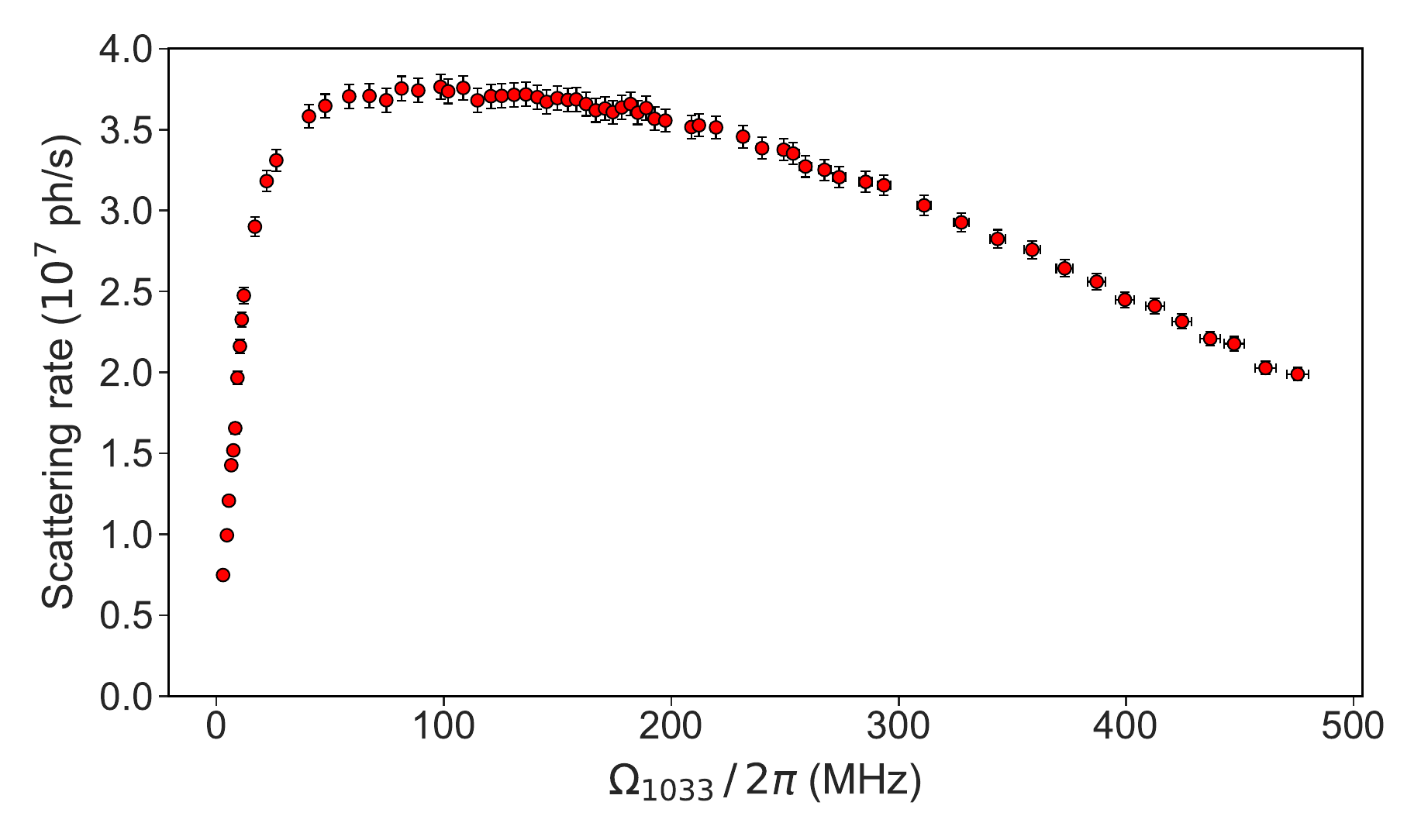}
    \caption{Scattering rate as a function of the 1033 nm Rabi frequency acquired with the following parameters : $\Delta_{422} = 0 \,\text{MHz}$, $\Delta_{1033} = -100 \,\text{MHz}$, $\Delta_{1003} = 0 \,\text{MHz}$, $\Omega_{422}/2\pi = 37\,\text{MHz}$, $\Omega_{1003}/2\pi = 200\,\text{MHz}$, $B \times \mu_B/h  = 3.6\,\text{MHz}$. \color{black}}
    \label{fig:compar1}
\end{figure}

In Fig. \ref{fig:compar1}, we show an intensity scan for which  $\Omega_{1033}/2\pi$ is varied from 10 to 475 MHz.

We find that the fluorescence intensity reaches a maximum for  $\Omega_{1033}/2\pi = 98 \, \textrm{MHz}$ and then decreases to lower values at higher laser intensities. 

In the next section, we compare this result with model calculations and discuss how closely this atomic ion system approximates a two-level atom.

\section{Discussion}
\label{Dis}

In the experiments we varied the repumping lasers intensities and kept constant the 422 nm probe laser intensity. 

\begin{figure}[h!]
    \centering
    \includegraphics[width=0.5\textwidth]{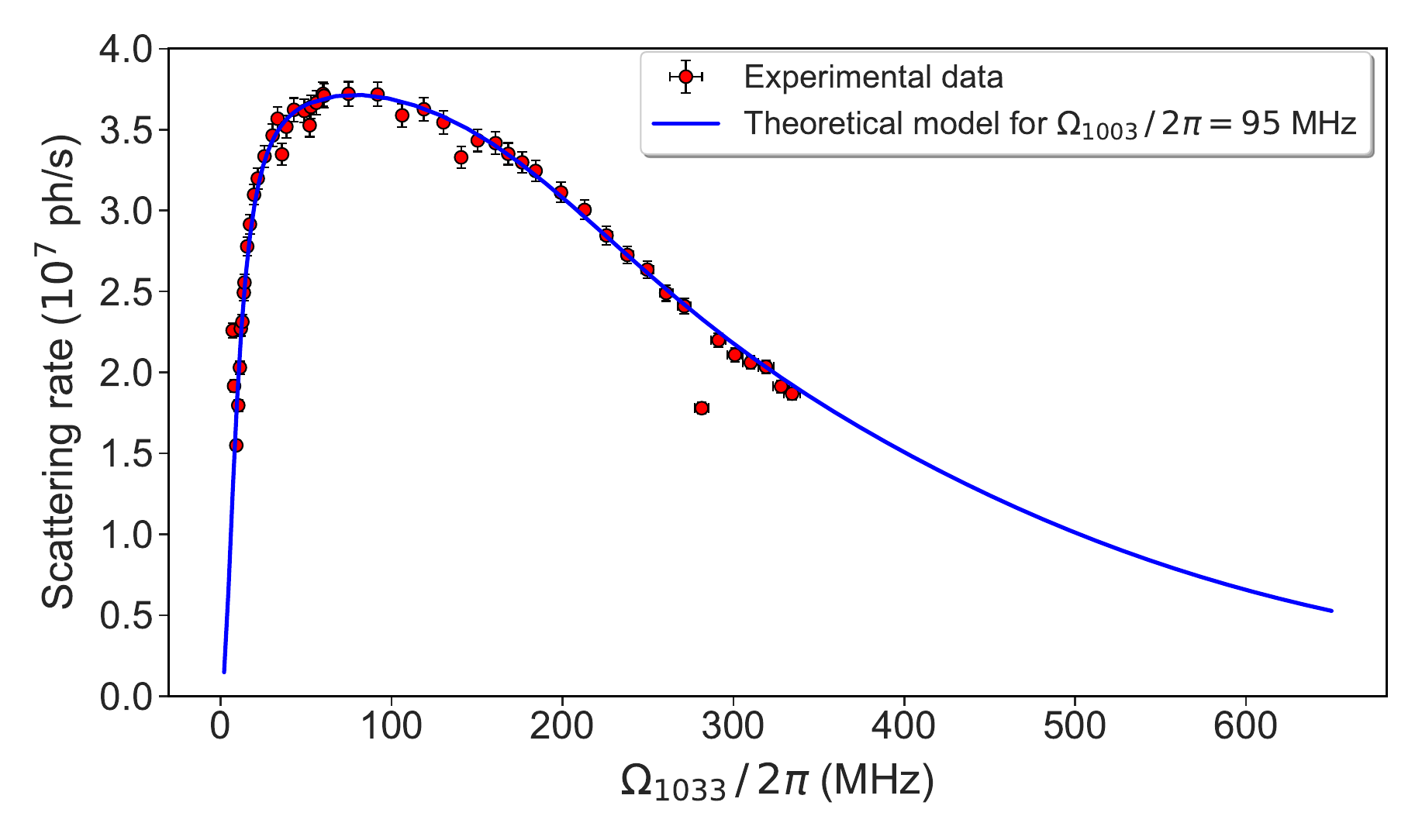}
    \caption{Scattering rate as a function of the 1033 nm Rabi frequency. The blue line corresponds to the above described model with the following parameters:   
    $\Delta_{422} = 0 \,\text{MHz}$,
    $\Delta_{1033} = -100 \,\text{MHz}$, $\Delta_{1003} = 0 \,\text{MHz}$, $\Omega_{422}/2\pi = 37\,\text{MHz}$, $\Omega_{1003}/2\pi = 95\,\text{MHz}$, $B \times \mu_B/h  = 3.6\,\text{MHz}$. \color{black}}
    \label{fig:img1}

    \vspace{1em} 

    \includegraphics[width=0.5\textwidth]{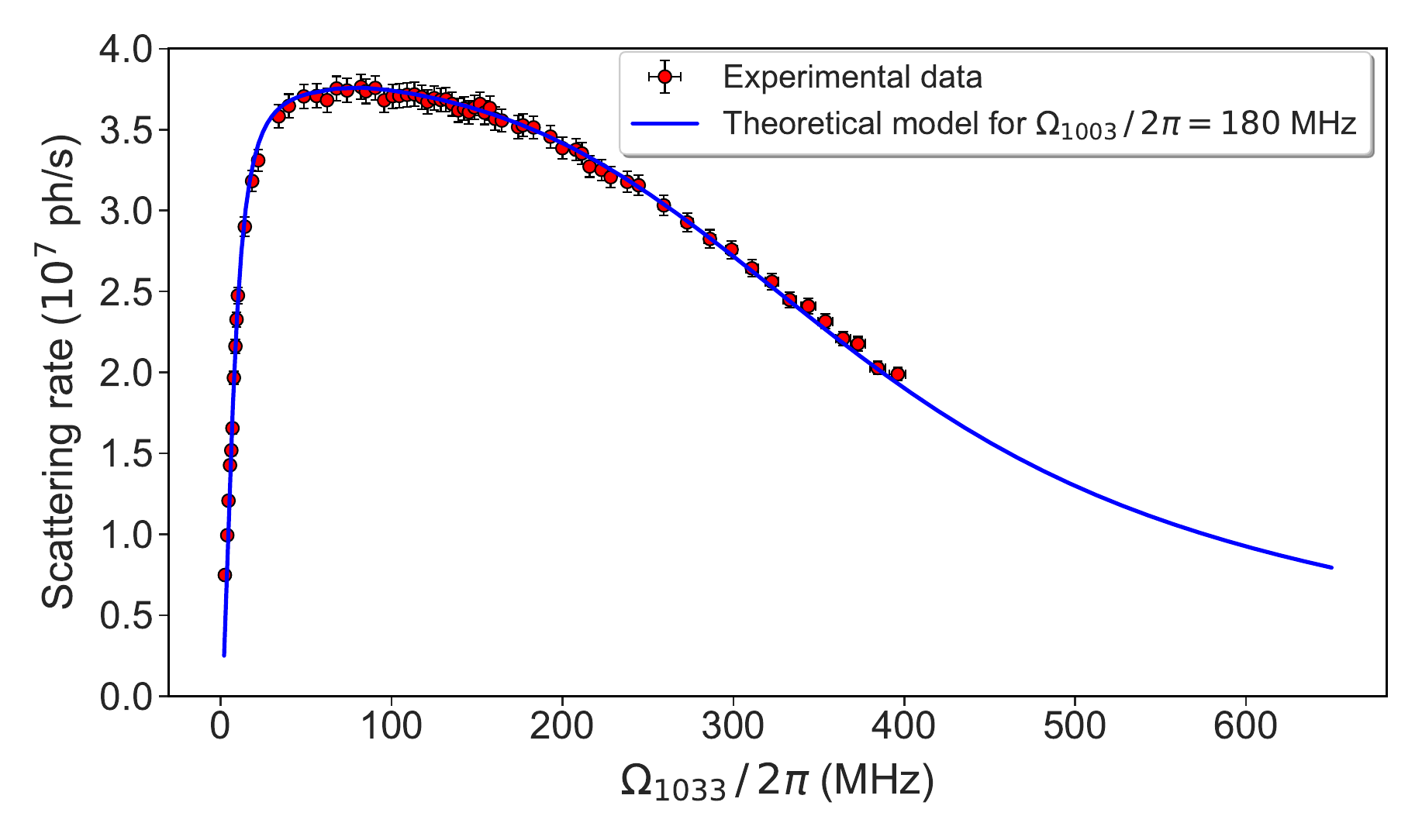}
    \caption{Scattering rate as a function of the 1033 nm Rabi frequency. The blue line corresponds to the above described model with the following parameters: $\Delta_{422} = 0 \,\text{MHz}$,
    $\Delta_{1033} = -100 \,\text{MHz}$, $\Delta_{1003} = 0 \,\text{MHz}$, $\Omega_{422}/2\pi = 37\,\text{MHz}$, $\Omega_{1003}/2\pi = 180\,\text{MHz}$, $B \times \mu_B/h  = 3.6\,\text{MHz}$. \color{black}}
    \label{fig:img2}
\end{figure}

In Figs. \ref{fig:img1} and \ref{fig:img2}, we compare experiments and calculations for two different intensities of the $1003$ nm laser. The corresponding calculations are materialized by the red dashed horizontal lines in Fig. \ref{fig:map2D}. The model catches all the features present in the experimental scans.

The $\Omega_{1033}/2\pi$ horizontal axes in Figs. \ref{fig:img1} and \ref{fig:img2} are set from the model. All the experimental  Rabi frequencies deduced from the measured beam size at the ion position and average power, are in agreement with the model within a factor $<1.5$.

Since the amplitude of the magnetic field at the ion position is determined by fitting fluorescence spectra such as in Fig. \ref{fig:spectrumIR1}, scattering rates are actually calculated without any adjustable parameters. 

\begin{figure}[H]
    \centering
    
    \includegraphics[width=0.45\textwidth]{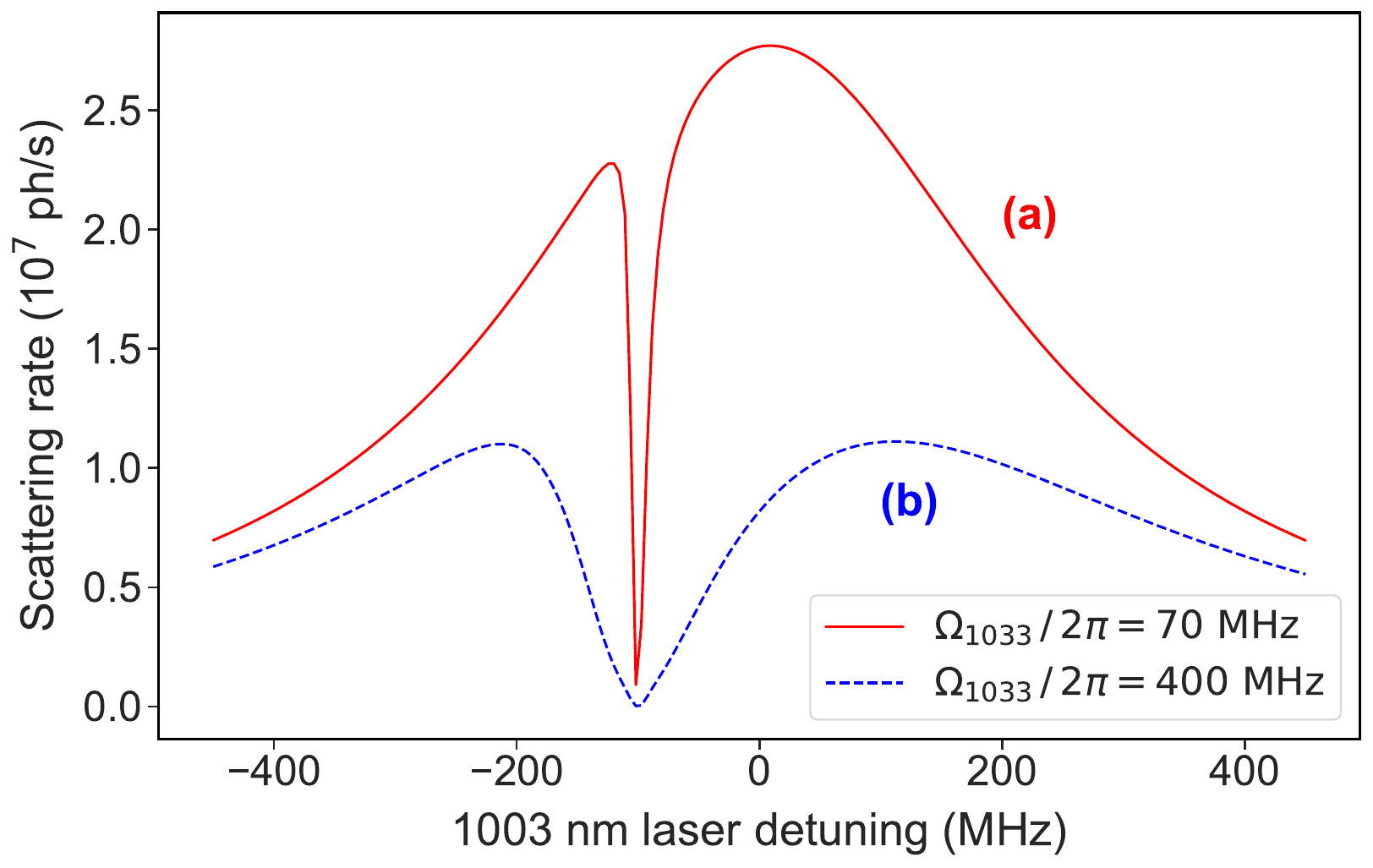}
    \caption{Scattering rate of the 422 nm photons as a function of the $1003$ nm laser frequency detuning $\Delta_{1003}$. The parameters used for all these calculations are as follows :  $\Delta_{1033} = -100 \,\text{MHz}$, $\Delta_{422} = 0 \,\text{MHz}$, $\Omega_{1003}/2\pi = 150\,\text{MHz}$, $\Omega_{422}/2\pi = 25\,\text{MHz}$. Calculations for $\Omega_{1033}/2\pi = 70\,\text{MHz}$ (a) (red curve) and for $\Omega_{1033}/2\pi = 400\,\text{MHz}$ (b) (blue dashed curve). The dip widens for larger laser intensities of the $1033$ nm laser. }
    \label{fig:ir2_spectrum}
\end{figure}

One question arises about the physical origin of the scattering rate decrease occurring for high ($\Omega/2\pi > 150\,\text{MHz}$) repumping fields intensities. The answer lies in the intensity-dependence of the coherent population trapping (CPT) in the $D_{3/2}-P_{3/2}-D_{5/2}$ $\Lambda$ system. To have a better insight of this phenomenon, we calculated the ion scattering rate at resonance as a function of  the $1003$ nm laser frequency detuning, presented in Fig. \ref{fig:ir2_spectrum}. 

A dip appears when the $1003$ nm and $1033$ nm laser fields have the same frequency detuning with respect to the $P_{3/2}$ state and is the signature in the fluorescence signal of the CPT in the $D$ states. In Fig. \ref{fig:ir2_spectrum},  we compare calculations for two different $1033$ nm laser intensities. At higher intensity for the $1033$ nm laser, the dip widens and the ion scattering rate is reduced.


In the three plots of Fig. \ref{fig:map2D},  the scattering rate exhibits a maximum for a given pair ($\Omega_{1003},\, \Omega_{1033}$). 
At the maximum of Fig. \ref{fig:map2D} (b), we calculate the total populations  in each of the five-level at saturation ($\Omega_{422}/2\pi > 10\,\text{GHz}$). We obtain 0.456, 0.456, 0.031, 0.032, 0.024 for the $P_{1/2}, S_{1/2}, D_{3/2},D_{5/2}, P_{3/2}$ states respectively. We note that the populations of  both $P_{1/2}$ and $S_{1/2}$ states, saturate at a value of 0.456 instead of 0.5  expected in the case of a two-level atom.

The residual population lying in the $D_{3/2}-P_{3/2}-D_{5/2}$ $\Lambda$ system comes from the broadening of the dark state in the mentioned $\Lambda$ system and to the finite lifetime of the $P_{3/2}$ state.

Let us note that while the total population of the $\Lambda$ system depends on the 422 nm Rabi frequency, the relative population distribution between the three states is independent of this parameter \cite{Tugaye:2019}.

\section{Conclusion}

In conclusion, we demonstrated that, in the case of the incoherent repumping scheme applied to a single $^{88}\text{Sr}^{+}$ ion, the simple analogy with a two-level atom is not sufficient to explain the experimental fluorescence levels. By combining a numerical model based on 18-sublevel optical Bloch equations with spectroscopic measurements, we have shown that the optimal repumping regime maximizing the scattering rate is achieved for a given combination of repumping laser intensities. This maximum scattering rate can only decrease when using more laser power, which may appear counterintuitive. 

These results provide a quantitative understanding of Doppler cooling atomic ions with metastable states and offer practical guidance for precision spectroscopy. They can be applied to other atomic species used in metrology, and quantum information experiments. These results also highlight that multi-level character remains essential even when the lineshape is Lorentzian and looks similar to that of a two-level atom.

\begin{acknowledgments}
We wish to acknowledge the support of M. Nicolas and B. Janvier for their technical assistance.
This study was partly funded by the french National Research Agency (project ESPRIT ANR-22-CE30-0028-03 and project HIT ANR-20-CE47-0013) and by Région Ile-de-France through the DIM Nano-k (DEQULOT grant).

\end{acknowledgments}

\section{Appendix}

\subsection{Modelisation}

\qquad We want to calculate the evolution of the density matrix of an ion driven by lasers and subjected to Zeeman splitting, under the approximation of the rotating wave. 

We thus have to put in equations three distinct phenomenon :

\begin{itemize}
\item Zeeman Splitting.
\item Stimulated absorption and emission.
\item Spontaneous emission.
\end{itemize}

\subsubsection{Zeeman Splitting.}

Each level of the ion is characterized by a set of two quantum numbers : 

\begin{itemize}
\item the total momentum : $J$
\item the orbital momentum : $l$
\end{itemize}

Those numbers are linked to two other quantities :

\begin{itemize}
\item the level degeneracy : $d = 2J+1$
\item the Landé g-factor : $g = 1+2(g_S-1)\dfrac{j-l}{2l+1}$ 
\end{itemize}

with $g_S = 2,0023$ the electron gyromagnetic ratio.
\medskip

The first gives the number of substates. A substate is designated by his magnetic number $m$ ranging from $-J$ to $J$ by steps of one. The Landé g-factor gives the amplitude of the Zeeman splitting under an external magnetic field, $B_{ext}$. In terms of energy and in term of frequency, the Zeeman shift of a substate $m$ is equal to :
\medskip
\begin{empheq}{align}
\Delta E_m &= g\mu_B B_{ext} \times m\\[3pt]
\Delta \nu_m &= g\dfrac{\mu_B}{h} B_{ext} \times m\\
\dfrac{\mu_B}{h} &= 14 \ {\rm GHz/T}
\end{empheq}
\medskip

With $\mu_B$ the Bohr magneton and $h$ the Planck constant.\\
In the proper basis, the Zeeman Hamiltonian elements are thus :
\begin{empheq}{align}
V_{Z,fi} = h\delta_{Z,fi} = \delta_{fi}\times h \ \! \Delta \nu_{f}
\end{empheq}

where $\delta_{fi}$ is the Kronecker symbol.

\subsubsection{Stimulated emission and absorption : }

The lasers are defined by three characteristics :

\begin{itemize}
\item Their frequency $\nu$ and which determine the pair(s) of states they couple : we neglect far off resonance excitations.
\item the amplitude of the light electric field at the position of the ion : $E_0$.
\item the polarisation of the beam and the wave vector : $\vec{\varepsilon}$ and $\vec{k}$.
\end{itemize}

If we neglect the spatial variations of $E_0$, the amplitude of the electric field is thus :
\begin{empheq}{align}
\vec{E}(\vec{x},t) = E_0\cos(2\pi\nu t- \vec{k}\vec{x})
\end{empheq}

The coherent evolution of the density matrix $\rho$ is thus given by :
\begin{empheq}{align}
i\hbar\dfrac{\partial \rho}{\partial t} & = [H,\rho] \label{evo_co}\\
H &= H_0 + V_Z + V
\end{empheq}

With $H_0$ the atomic Hamiltonian and $V$ representing the interaction with the radiation. For two non degenerate levels $a$ and $b$ we have in the electric dipole approximation :
\begin{empheq}{align}
V_{ab} &= -\bra{a}\vec{d}.\vec{\varepsilon}\ket{b} E_0 \cos(2\pi\nu t) = -h\Omega_{ab} \cos(2\pi\nu t)
\end{empheq}

With $\Omega_{ab}$ the Rabi frequency of the transition. 
Note that it is proportional to the field amplitude, and thus to the squareroot of the intensity. It also depends on beam alignment and shape.
\medskip

If we write the equations in the interaction representation toward $H_0$ :
\begin{empheq}{align}
s_{ab} = \rho_{ab}\exp\left (-\dfrac{2i\pi}{h}(E_a-E_b)t\right )
\end{empheq}
 
we obtain after taking the rotating wave approximation :


\begin{empheq}{align}
\begin{split}
\dot{s}_{ab} ={}&\; i\pi\sum_k \Omega_{ak}s_{kb}e^{2i\pi\Delta_{ak}t}
               - s_{ak}\Omega_{kb}e^{2i\pi\Delta_{kb}t} \\
                &\; -2i\pi(\Delta\nu_a-\Delta\nu_b)s_{ab}
\end{split}
\end{empheq}

Where we have defined the detuning in frequency units (not pulsation) : 

\begin{empheq}{align}
\Delta_{ab} &= \dfrac{E_a-E_b}{h}-\nu_{ab}\\
\nu_{ab}&\times(E_a-E_b) > 0, \quad |\nu_{ab}| = \nu
\end{empheq}
\medskip

With this sign convention, the detuning is positive when the laser is shifted to the blue of the resonance.

To completely eliminate the time dependences we have to define detuning for states coupled through multi photon processes :
\begin{empheq}{align}
\Delta_{ab} &= \Delta_{ak}+\Delta_{kb}
\end{empheq}
\medskip

This is not always possible as it impose constraints on lasers frequencies when two different path couple the same states.
\medskip

We then write :
\begin{empheq}{align}
\sigma_{ab} = s_{ab}\exp\left(2i\pi\Delta_{ab}t\right)
\end{empheq}

And thus obtain :
\begin{empheq}{align}
\begin{split}
\frac{1}{2\pi}\,\dot{\sigma}_{ab}
   ={}&\; \frac{i}{2}\sum_k\!\left(\Omega_{ak}\sigma_{kb}
            - \sigma_{ak}\Omega_{kb}\right) \\
      &\; -\, i\left(\Delta_{ab} + \Delta\nu_a - \Delta\nu_b\right)\sigma_{ab}
\end{split}
\end{empheq}

Writing $*$ the elementwise product of two matrices this simplifies into :

\begin{empheq}{align}
\frac{1}{2\pi} \ \dot{\sigma} = \dfrac{i}{2}[\Omega+\delta_Z,\sigma] - i\Delta*\sigma
\end{empheq}

\paragraph{$\blacktriangleright$ sublevels modification of Rabi frequencies :}

For two degenerate levels $a,J'$ and $b,J$ the polarisation of the light defines how the sublevels are coupled. For two substates m' and m we have~:

\begin{empheq}{alignat = 2}
\Omega_{a,b}^{m',m} &= \dfrac{\mu_{ab}E_0}{h} &\sum_{q = -1,0,1} C^{J,m}_{J',m',1,q}\times \varepsilon_q\\
& = \ \ \ \! \Omega_{ab} &\sum_{q = -1,0,1} C^{J,m}_{J',m',1,q}\times \varepsilon_q
\end{empheq}
\medskip

Where $\mu_{ab}$ only depends on the levels $a$ and $b$ and :

\begin{empheq}{align}
\vec{e}_0 &= \vec{e}_z, \ \vec{e}_{\pm 1} = \mp (\vec{e}_x \pm i\vec{e}_y)/\sqrt{2}\\
\varepsilon_q &= \vec{\varepsilon}.\vec{e}_q\ ,\qquad q = -1,0,1
\end{empheq}
\medskip

The vectors $\vec{e}_q$ are associated to the three polarisation of the beam according to the magnetic field :

\begin{itemize}
\item $\sigma^\pm$ for $\vec{e}_{\pm 1}$, the two circular polarisation for a beam collinear to the magnetic field.
\item $\pi$ for $\vec{e}_{0}$, the linear polarisation parallel to the magnetic field.
\end{itemize}

And :
\begin{empheq}{align}
C^{J,m}_{J',m',1,q} &=\vert{J',m',1,q}J,m\ \!\rangle
\end{empheq}
\medskip

The $C^{J,m}_{J',m',1,q}$ are the well-known Clebsh-Gordan coefficients. A table is given in appendix • . For $C^{J,m}_{J',m',1,q}$ to be not equal to zero, the conservation of the light and ion's total angular momentum projection on the quantification axis has to hold. This mean that $m$ has to be equal to $m'+q$, thus at most one term of the sum counts.

\paragraph{$\blacktriangleright$ lasers linewidth :}

Lasers linewidth are dealed with in an empirical fashion. Indeed for small laser linewidth compared to the transition linewidth, adding a real part $\gamma_{ab}$ to the detuning $\Delta_{ab}$, $\Delta_{ba}$ corresponding to coupling between $a$ and $b$ emulate decoherence due to external factors like laser linewidth or magnetic field fluctuation :

\begin{empheq}{align}
\dfrac{1}{2\pi}\dot{\sigma} &= \dfrac{i}{2}\left [\Omega + \delta_Z,\sigma\right ] - (\gamma + i\Delta)*\sigma 
\end{empheq}

\subsubsection{Spontaneous emission}

As long as we only care about the mean level of fluorescence of an ensemble of ions, and not the details of the emission time, spontaneous emission is described by the Linblad operator :

\medskip

\begin{empheq}{align}
\frac{1}{2\pi}\dot{\sigma}
   &= \frac{i}{2}\,[\Omega+\delta_Z,\sigma]
      - (\gamma + i\Delta)*\sigma
      + L[\sigma]
\end{empheq}

\begin{empheq}{align}
L[\sigma]
   &= \sum_{f,i}\frac{1}{2}\!\left(
         C_{fi}^\dag C_{fi}\sigma
       + \sigma C_{fi}^\dag C_{fi}
     \right)
      + C_{fi}\sigma C_{fi}^\dag
\end{empheq}

\begin{empheq}{align}
C_{fi}
   &= \sqrt{\Gamma_{fi}}\ket{f}\bra{i}
\end{empheq}

Where each $C_{fi}$ is associated with a decay from the state $i$ to $f$. The coefficients $\Gamma_{fi}$ are the decay frequencies. For an electron in the state $i$, the probability to decay to the state $i$ in a window d$t$ is equal to :
\begin{empheq}{align}
p_{i \rightarrow f} = 2\pi\Gamma_{fi}\ \!{\rm d}t
\end{empheq}
\medskip

There is two type of decay that are of importance for our experiments : the electric dipole decay, which matters for the $P$ states and the electric quadrupole decay which matters for the forbidden transitions $S$ to $D$.
\medskip

For degenerated levels, the transition probability between two substates is a bit modified. For an initial state $\ket{i,J,m}$ and a final state $\ket{f,J',m'}$ we have :
%
\medskip

\begin{empheq}{align}
\begin{aligned}
\Gamma_{fi}^{m',m} &= \Gamma_{fi} \sum_{q=-1}^1 {C^{J,m}_{J',m',1,q}}^2 \\
&\makebox[5cm][l]{for electric dipole transitions}
\end{aligned} \\[6pt]
\begin{aligned}
\Gamma_{fi}^{m',m} &= \Gamma_{fi}^2 \sum_{q=-2}^2 {C^{J,m}_{J',m',2,q}}^2 \\
&\makebox[5cm][l]{for electric quadrupole transitions}
\end{aligned}
\end{empheq}

If a transition is allowed :

\begin{empheq}{align}
\sum_{q,m'} {C^{J,m}_{J',m',k,q}}^2 &= 1
\end{empheq}

Thus the lifetimes of each substates are equal.

\bibliographystyle{unsrt}
\bibliography{biblio_total_2024_modjpl}

@article{Tugaye:2019,
	author = {Tugay\'e, Vincent and Likforman, Jean-Pierre and Guibal, Samuel and Guidoni, Luca},
	doi = {10.1103/PhysRevA.99.023412},
	issue = {2},
	journal = {Phys. Rev. A},
	month = {Feb},
	numpages = {7},
	pages = {023412},
	publisher = {American Physical Society},
	title = {Absolute single-ion thermometry},
	url = {https://link.aps.org/doi/10.1103/PhysRevA.99.023412},
	volume = {99},
	year = {2019}}

@article{Ludlow:2015,
	author = {Ludlow, Andrew D. and Boyd, Martin M. and Ye, Jun and Peik, E. and Schmidt, P. O.},
	doi = {10.1103/RevModPhys.87.637},
	issue = {2},
	journal = {Rev. Mod. Phys.},
	month = {Jun},
	numpages = {65},
	pages = {637--701},
	publisher = {American Physical Society},
	title = {Optical atomic clocks},
	url = {https://link.aps.org/doi/10.1103/RevModPhys.87.637},
	volume = {87},
	year = {2015}}

@article{Sikorsky:2017,
	author = {Sikorsky, Tomas and Meir, Ziv and Akerman, Nitzan and Ben-shlomi, Ruti and Ozeri, Roee},
	doi = {10.1103/PhysRevA.96.012519},
	issue = {1},
	journal = {Phys. Rev. A},
	month = {Jul},
	numpages = {10},
	pages = {012519},
	publisher = {American Physical Society},
	title = {Doppler cooling thermometry of a multilevel ion in the presence of micromotion},
	url = {https://link.aps.org/doi/10.1103/PhysRevA.96.012519},
	volume = {96},
	year = {2017}}

@article{Morigi:2000,
	author = {Morigi, Giovanna and Eschner, J\"urgen and Keitel, Christoph H.},
	doi = {10.1103/PhysRevLett.85.4458},
	issue = {21},
	journal = {Phys. Rev. Lett.},
	month = {Nov},
	numpages = {0},
	pages = {4458--4461},
	publisher = {American Physical Society},
	title = {Ground State Laser Cooling Using Electromagnetically Induced Transparency},
	url = {https://link.aps.org/doi/10.1103/PhysRevLett.85.4458},
	volume = {85},
	year = {2000}}

@article{Janacek:2018,
	author = {Janacek, H. and Steane, A. M. and Lucas, D. M. and Stacey, D. N.},
	booktitle = {Journal of Modern Optics},
	da = {2018/03/30},
	date = {2018/03/30},
	doi = {10.1080/09500340.2018.1428771},
	isbn = {0950-0340},
	journal = {Journal of Modern Optics},
	journal1 = {Journal of Modern Optics},
	m3 = {doi: 10.1080/09500340.2018.1428771},
	month = {03},
	number = {5-6},
	pages = {577--584},
	publisher = {Taylor \& Francis},
	title = {The effect of atomic response time in the theory of Doppler cooling of trapped ions},
	ty = {JOUR},
	url = {https://doi.org/10.1080/09500340.2018.1428771},
	volume = {65},
	year = {2018},
	year1 = {2018}}

@article{Arimondo:1976,
	author = {Arimondo, E. and Orriols, G.},
	da = {1976/11/01},
	doi = {10.1007/BF02746514},
	id = {Arimondo1976},
	isbn = {1827-613X},
	journal = {Lettere al Nuovo Cimento (1971-1985)},
	number = {10},
	pages = {333--338},
	title = {Nonabsorbing atomic coherences by coherent two-photon transitions in a three-level optical pumping},
	ty = {JOUR},
	url = {https://doi.org/10.1007/BF02746514},
	volume = {17},
	year = {1976}}

@article{Likforman:2016,
	author = {Likforman, Jean-Pierre and Tugay\'e, Vincent and Guibal, Samuel and Guidoni, Luca},
	doi = {10.1103/PhysRevA.93.052507},
	issue = {5},
	journal = {Phys. Rev. A},
	month = {May},
	numpages = {9},
	pages = {052507},
	publisher = {American Physical Society},
	title = {Precision measurement of the branching fractions of the $5p\phantom{\rule{0.16em}{0ex}}^{2}P_{1/2}$ state in $^{88}\mathrm{Sr}^{+}$ with a single ion in a microfabricated surface trap},
	url = {http://link.aps.org/doi/10.1103/PhysRevA.93.052507},
	volume = {93},
	year = {2016}}

@book{Grynberg:2010,
	author = {Gilbert Grynberg and Alain Aspect and Claude Fabre},
	doi = {isbn: 9780521551120},
	edition = {1rst},
	publisher = {Cambridge University Press},
	title = {Introduction to Quantum Optics: From the Semi-classical Approach to Quantized Light},
	year = {2010}}

@article{Biemont:2000,
	author = {E. Bi{\'e}mont and J. Lidberg and S. Mannervik and L.-O. Norlin and P. Royen and A. Schmitt and W. Shi and X. Tordoir},
	doi = {10.1007/s100530070063},
	journal = {Eur. Phys. J. D},
	number = 3,
	pages = {355-365},
	title = {Lifetimes of metastable states in Sr II},
	url = {http://dx.doi.org/10.1007/s100530070063},
	volume = 11,
	year = 2000}

@article{Ramm:2013,
	author = {Ramm, Michael and Pruttivarasin, Thaned and Kokish, Mark and Talukdar, Ishan and H{\"a}ffner, Hartmut},
	doi = {10.1103/PhysRevLett.111.023004},
	issue = {2},
	journal = {Phys. Rev. Lett.},
	month = {Jul},
	numpages = {4},
	pages = {023004},
	publisher = {American Physical Society},
	title = {Precision Measurement Method for Branching Fractions of Excited ${P}_{1/2}$ States Applied to $^{40}\mathrm{Ca}^{\mathbf{+}}$},
	url = {http://link.aps.org/doi/10.1103/PhysRevLett.111.023004},
	volume = {111},
	year = {2013}}

@article{Gardner:2014,
	author = {Gardner, Amy and Sheridan, Kevin and Groom, William and Seymour-Smith, Nicolas and Keller, Matthias},
	booktitle = {Applied Physics B},
	da = {2014/11/01},
	doi = {10.1007/s00340-014-5891-1},
	isbn = {0946-2171},
	j2 = {Appl. Phys. B},
	journal = {Appl. Phys. B},
	la = {English},
	number = {2},
	pages = {755-762},
	publisher = {Springer Berlin Heidelberg},
	title = {Precision spectroscopy technique for dipole-allowed transitions in laser-cooled ions},
	ty = {JOUR},
	url = {http://dx.doi.org/10.1007/s00340-014-5891-1},
	volume = {117},
	year = {2014}}

@article{Meir:2014,
	author = {Meir, Z. and Schwartz, O. and Shahmoon, E. and Oron, D. and Ozeri, R.},
	doi = {10.1103/PhysRevLett.113.193002},
	issue = {19},
	journal = {Phys. Rev. Lett.},
	month = {Nov},
	numpages = {5},
	pages = {193002},
	publisher = {American Physical Society},
	title = {Cooperative Lamb Shift in a Mesoscopic Atomic Array},
	url = {http://link.aps.org/doi/10.1103/PhysRevLett.113.193002},
	volume = {113},
	year = {2014}}

@article{Hansch:1975,
	author = {T.W. H{\"a}nsch and A.L. Schawlow},
	doi = {http://dx.doi.org/10.1016/0030-4018(75)90159-5},
	issn = {0030-4018},
	journal = {Optics Communications},
	number = {1},
	pages = {68 - 69},
	title = {Cooling of gases by laser radiation},
	url = {http://www.sciencedirect.com/science/article/pii/0030401875901595},
	volume = {13},
	year = {1975}}

@article{Wineland:1975a,
	author = {D. Wineland and H. Dehmelt},
	journal = {Bull. Am. Phys. Soc.},
	pages = {637},
	title = {Proposed 10$^14$ $\Delta\nu<\nu$ Laser Fluorescence Spectroscopy on Tl$^+$ Mono-Ion Oscillator III},
	volume = {20},
	year = {1975}}

@book{Ghosh:1995,
	author = {P. Ghosh},
	publisher = {Oxford Science Publications},
	title = {Ion Traps},
	year = {1995}}

@article{Dehmelt:1967,
	author = {H. G. Dehmelt},
	journal = {Advances in Atomic and Molecular Physics},
	pages = {53 -72},
	publisher = {Academic Press, New York},
	title = {Radiofrequency spectroscopy of stored ions I : storage},
	volume = {3},
	year = {1967}}

@article{Cirac:1995,
	author = {Cirac, J. I. and Zoller, P.},
	doi = {10.1103/PhysRevLett.74.4091},
	issue = {20},
	journal = {Phys. Rev. Lett.},
	month = {May},
	pages = {4091--4094},
	publisher = {American Physical Society},
	title = {Quantum Computations with Cold Trapped Ions},
	url = {http://link.aps.org/doi/10.1103/PhysRevLett.74.4091},
	volume = {74},
	year = {1995}}

@article{Pinnington:1995,
	author = {E H Pinnington and R W Berends and M Lumsden},
	journal = {J. Phys. B: At. Mol. Opt. Phys.},
	number = {11},
	pages = {2095},
	title = {Studies of laser-induced fluorescence in fast beams of Sr + and Ba + ions},
	url = {http://stacks.iop.org/0953-4075/28/i=11/a=009},
	volume = {28},
	year = {1995}}

@article{Paul:1990,
	author = {Paul, Wolfgang},
	doi = {10.1103/RevModPhys.62.531},
	journal = {Rev. Mod. Phys.},
	month = {Jul},
	number = {3},
	numpages = {9},
	pages = {531--540},
	publisher = {American Physical Society},
	title = {Electromagnetic traps for charged and neutral particles},
	volume = {62},
	year = {1990}}

@article{Allcock:2010,
	author = {D. T. C. Allcock and J. A. Sherman and D. N. Stacey and A. H. Burrell and M. J. Curtis and G. Imreh and N. M. Linke and D. J. Szwer and S. C. Webster and A. M. Steane and D. M. Lucas},
	journal = {New J. Phys.},
	number = {5},
	pages = {053026},
	title = {Implementation of a symmetric surface-electrode ion trap with field compensation using a modulated {R}aman effect},
	url = {http://stacks.iop.org/1367-2630/12/i=5/a=053026},
	volume = {12},
	year = {2010}}

@article{Wesenberg:2007,
	author = {Wesenberg, J. H. and Epstein, R. J. and Leibfried, D. and Blakestad, R. B. and Britton, J. and Home, J. P. and Itano, W. M. and Jost, J. D. and Knill, E. and Langer, C. and Ozeri, R. and Seidelin, S. and Wineland, D. J.},
	doi = {10.1103/PhysRevA.76.053416},
	journal = {Phys. Rev. A},
	month = {Nov},
	number = {5},
	numpages = {11},
	pages = {053416},
	publisher = {American Physical Society},
	title = {Fluorescence during Doppler cooling of a single trapped atom},
	volume = {76},
	year = {2007}}

@article{Seidelin:2006,
	author = {Seidelin, S. and Chiaverini, J. and Reichle, R. and Bollinger, J. J. and Leibfried, D. and Britton, J. and Wesenberg, J. H. and Blakestad, R. B. and Epstein, R. J. and Hume, D. B. and Itano, W. M. and Jost, J. D. and Langer, C. and Ozeri, R. and Shiga, N. and Wineland, D. J.},
	doi = {10.1103/PhysRevLett.96.253003},
	journal = {Phys. Rev. Lett.},
	month = {Jun},
	number = {25},
	numpages = {4},
	pages = {253003},
	publisher = {American Physical Society},
	title = {Microfabricated Surface-Electrode Ion Trap for Scalable Quantum Information Processing},
	volume = {96},
	year = {2006}}

@article{Chiaverini:2005,
	author = {J. Chiaverini and R. B. Blakestad and J. Britton and J. D. Jost and C. Langer and D. Leibfried and R. Ozeri and D. J. Wineland},
	citeseerurl = {http://arxiv.org/abs/quant-ph/0501147},
	journal = {Quantum Inf. Comput.},
	number = {6},
	pages = {419-439},
	title = {Surface-Electrode Architecture for Ion-Trap Quantum Information Processing},
	volume = {5},
	year = {2005}}

@Book{Foot:2004,
  author    = {C. J. Foot},
  editor    = {Oxford University Press},
  publisher = {Oxford University Press},
  title     = {Atomic Physics},
  year      = {2004},
}

@Article{Arecchi:1965,
  author    = {Arecchi, F. and Bonifacio, R.},
  journal   = {IEEE Journal of Quantum Electronics},
  title     = {Theory of optical maser amplifiers},
  year      = {1965},
  issn      = {0018-9197},
  month     = jul,
  number    = {4},
  pages     = {169--178},
  volume    = {1},
  doi       = {https://doi.org/10.1109/JQE.1965.1072212},
  publisher = {Institute of Electrical and Electronics Engineers (IEEE)},
}

@Book{Allen:1975,
  author = {A. Allen, J.H. Eberly},
  title  = {Optical resonance and two-level atoms},
  year   = {1975},
}

@PhdThesis{Tugaye:2020,
  author = {Vincent Tugaye},
  school = {Universite Paris Cite},
  title  = {Spectroscopie et thermometrie d ions uniques 88Sr+ captures dans des micropieges surfaciques aspects theoriques et experimentaux},
  year   = {2020},
}

@PhdThesis{Janacek:2015,
  author = {Hugo Alexander Janacek},
  school = {Oxford},
  title  = {Optical Bloch equations for simulatingtrapped-ion qubits},
  year   = {2015},
}

@Article{Maguire:2006,
  author    = {Maguire, L P and Bijnen, R M W van and Mese, E and Scholten, R E},
  journal   = {Journal of Physics B: Atomic, Molecular and Optical Physics},
  title     = {Theoretical calculation of saturated absorption spectra for multi-level atoms},
  year      = {2006},
  issn      = {1361-6455},
  month     = may,
  number    = {12},
  pages     = {2709--2720},
  volume    = {39},
  doi       = {doi:10.1088/0953-4075/39/12/007},
  publisher = {IOP Publishing},
}

@techreport{NIST:2025,
  author        = {National Institute of Standards and Technology},
  title         = {NIST Atomic Spectra Database},
  institution   = {U.S. Department of Commerce},
  address       = {Gaithersburg, MD},
  year          = {2025},
  month         = {11},
  type          = {},
  number        = {NIST Standard Reference Database 78},
  DOI           = {10.18434/T4W30F},
  note          = {https://physics.nist.gov/asd}
}

@Article{Marshall:2025,
  author    = {Marshall, Mason C. and Castillo, Daniel A. Rodriguez and Arthur-Dworschack, Willa J. and Aeppli, Alexander and Kim, Kyungtae and Lee, Dahyeon and Warfield, William and Hinrichs, Joost and Nardelli, Nicholas V. and Fortier, Tara M. and Ye, Jun and Leibrandt, David R. and Hume, David B.},
  title     = {High-Stability Single-Ion Clock with 5.5 x 10-19 Systematic Uncertainty},
  year      = {2025},
  copyright = {arXiv.org perpetual, non-exclusive license},
  doi       = {https://doi.org/10.48550/arXiv.2504.13071},
  publisher = {arXiv},
}

@Article{Thompson:1990,
  author    = {Thompson, R C},
  journal   = {Measurement Science and Technology},
  title     = {Precision measurement aspects of ion traps},
  year      = {1990},
  issn      = {1361-6501},
  month     = feb,
  number    = {2},
  pages     = {93--105},
  volume    = {1},
  doi       = {10.1088/0957-0233/1/2/001},
  publisher = {IOP Publishing},
}

@Article{FossFeig:2025,
  author    = {Foss-Feig, Michael and Pagano, Guido and Potter, Andrew C. and Yao, Norman Y.},
  journal   = {Annual Review of Condensed Matter Physics},
  title     = {Progress in Trapped-Ion Quantum Simulation},
  year      = {2025},
  issn      = {1947-5462},
  month     = mar,
  number    = {1},
  pages     = {145--172},
  volume    = {16},
  doi       = {https://doi.org/10.1146/annurev-conmatphys-032822-045619},
  publisher = {Annual Reviews},
}

@Article{Bruze:2019,
  author    = {Bruzewicz, Colin D. and Chiaverini, John and McConnell, Robert and Sage, Jeremy M.},
  journal   = {Applied Physics Reviews},
  title     = {Trapped-ion quantum computing: Progress and challenges},
  year      = {2019},
  issn      = {1931-9401},
  month     = may,
  number    = {2},
  volume    = {6},
  doi       = {https://doi.org/10.1063/1.5088164},
  publisher = {AIP Publishing},
}

@article{Hughes:2025,
  author    = {Hughes, A. C. and Srinivas, R. and L{\"o}schnauer, C. M. and Knaack, H. M. and Matt, R. and Ballance, C. J. and Malinowski, M. and Harty, T. P. and Sutherland, R. T.},
  title     = {Trapped-ion two-qubit gates with $>$99.99\% fidelity without ground-state cooling},
  journal   = {arXiv preprint arXiv:2510.17286},
  year      = {2025},
  note      = {\url{https://arxiv.org/abs/2510.17286}}
}

\end{document}